PHYSICS

# Inducing ferromagnetism and Kondo effect in platinum by paramagnetic ionic gating

Lei Liang, Qihong Chen, Jianming Lu,* Wytse Talsma, Juan Shan, Graeme R. Blake, Thomas T. M. Palstra,[†] Jianting Ye[‡]

Electrically controllable magnetism, which requires the field-effect manipulation of both charge and spin degrees of freedom, has attracted growing interest since the emergence of spintronics. We report the reversible electrical switching of ferromagnetic (FM) states in platinum (Pt) thin films by introducing paramagnetic ionic liquid (PIL) as the gating media. The paramagnetic ionic gating controls the movement of ions with magnetic moments, which induces itinerant ferromagnetism on the surface of Pt films, with large coercivity and perpendicular anisotropy mimicking the ideal two-dimensional Ising-type FM state. The electrical transport of the induced FM state shows Kondo effect at low temperature, suggesting spatially separated coexistence of Kondo scattering beneath the FM interface. The tunable FM state indicates that paramagnetic ionic gating could serve as a versatile method to induce rich transport phenomena combining field effect and magnetism at PIL-gated interfaces.

## INTRODUCTION

The ongoing quest to control magnetism by an electric field has attracted growing interest in both fundamental sciences and technological applications (1–5). In diluted magnetic semiconductors, switching magnetization can be achieved by modifying the density and type of carriers with external electric field (1, 5–7). In multiferroic materials, the electric polarization can couple with the magnetization due to exchange striction effects (8–10). However, both aforementioned approaches require a strong electric field and usually reach magnetic ordering below room temperature, making them less feasible for applications. The materials showing high Curie temperatures ($T_C$) are generally metallic, which is difficult to manipulate by the field effect due to their intrinsically large carrier densities and, consequently, their short Thomas-Fermi screening lengths. The application of ionic liquids (ILs) on gating (Fig. 1A) has achieved inducing quantum phase transitions in many insulators (11–13) and semiconductors (14, 15). A large number of carriers accumulated by ionic gating can even tune metallic devices (2, 16–19). However, so far, ionic gating can only gradually vary metallic ferromagnetic (FM) materials (2, 3, 17), without realizing marked changes, such as ON/OFF switching of FM states.

Physically, the Stoner model of band ferromagnets explains spontaneous spin splitting in FM metals such as Fe, Co, and Ni, requiring the product of the density of states (DOS) at Fermi energy $\rho_F$ and the exchange integral $I$ larger than unity. Platinum (Pt) is normally regarded as an exchange-enhanced paramagnetic (PM) metal on the verge of FM instability. Hence, applying an electric field could induce the FM state in Pt if the enhanced product $I\rho_F$ satisfies the Stoner criterion, which might subsequently evoke marked changes in both magnetism and electrical transport. Meanwhile, decreasing the coordination number of the nearest neighboring atoms at the surface results in reduced electronic bandwidth (20). Consequently, ferromagnetism can be induced when the product of $I$ and $\rho_F$ is strongly enhanced by reducing dimensionality (21). For example, although not electrically controllable, the isolated Pt nanoparticles show ferromagnetism if their surfaces are perturbed by chemisorption (22).

Despite the fact that ionic gating is capable of tuning a large number of carriers, it is highly demanding to realize the field-effect control of the spin degree of freedom. Here, paramagnetic ionic liquids (PILs), composed of anions containing transition metal with unpaired d orbitals, are introduced as gating media to induce magnetic interactions at the gated channel surface (Fig. 1B and see the Supplementary Materials). Therefore, it extends the conventional ionic gating to the spin tunability: the second intrinsic characteristic of the electron. Replacing the organic anions with metal complexes maintains the general physicochemical properties of ILs, such as low melting temperature ($T_m$ < 200 K), negligible vapor pressure, and a large electrochemical window. Guided by these prerequisites that are crucial for ionic gating, butylmethylimidazolium tetrachloroferrate (BMIM[FeCl$_4$]) was synthesized for all experiments discussed below. All five 3d orbitals of Fe$^{3+}$ in the anions are unpaired, giving a total spin quantum number of $S$ = 5/2 (high-spin state). This PIL responds actively to external magnetic field even at room temperature (fig. S1). Magnetic susceptibility measurement shows that the PM BMIM[FeCl$_4$] has a large effective moment μ = 5.87 μ$_B$, following Curie's law down to 2 K (fig. S2).

## RESULTS

### Gate-controllable electrical transport and field-induced ferromagnetism in Pt

Pt thin films with various thicknesses were prepared by magnetron sputtering. The films were patterned into Hall bar geometries and gated by BMIM[FeCl$_4$], as shown schematically in Fig. 1 (A to C). The gate voltage $V_G$ dependence of the sheet resistance $R_s$ (device A, thickness $t$ = 8.0 nm) shows that the $R_s$ can be reversibly controlled with negligible leak current $I_G$ (Fig. 1D). The electrostatic nature of gating is further confirmed by a chronoamperometry experiment (fig. S3). According to ab initio calculation, the DOS peak of Pt lies slightly below the Fermi level ($E_F$) (23). Therefore, we expect that depleting carriers with negative $V_G$ by driving the magnetic anions toward the Pt surface might satisfy the Stoner criterion; it would do this by increasing both $\rho_F$ and $I$, due to the possible d-d interaction between Pt and FeCl$_4^-$. However, the negatively gated Pt maintains the PM state despite the increase of $|R_{xy}/B|$ (fig. S4). In contrast, FM states can be induced when the $V_G$ dependence of the $R_s$ shows an obvious drop at $V_G$ > 0 (Fig. 1D), evidenced by the anomalous Hall effect (AHE) with clear hysteresis (Fig. 1E). This finding

Zernike Institute for Advanced Materials, University of Groningen, Nijenborgh 4, 9747 AG Groningen, Netherlands.
*Present address: State key laboratory for mesoscopic physics, Peking University, Beijing 100871, People's Republic of China.
†Present address: College van Bestuur, Universiteit Twente, 7500 AE Enschede, Netherlands.
‡Corresponding author. Email: j.ye@rug.nl

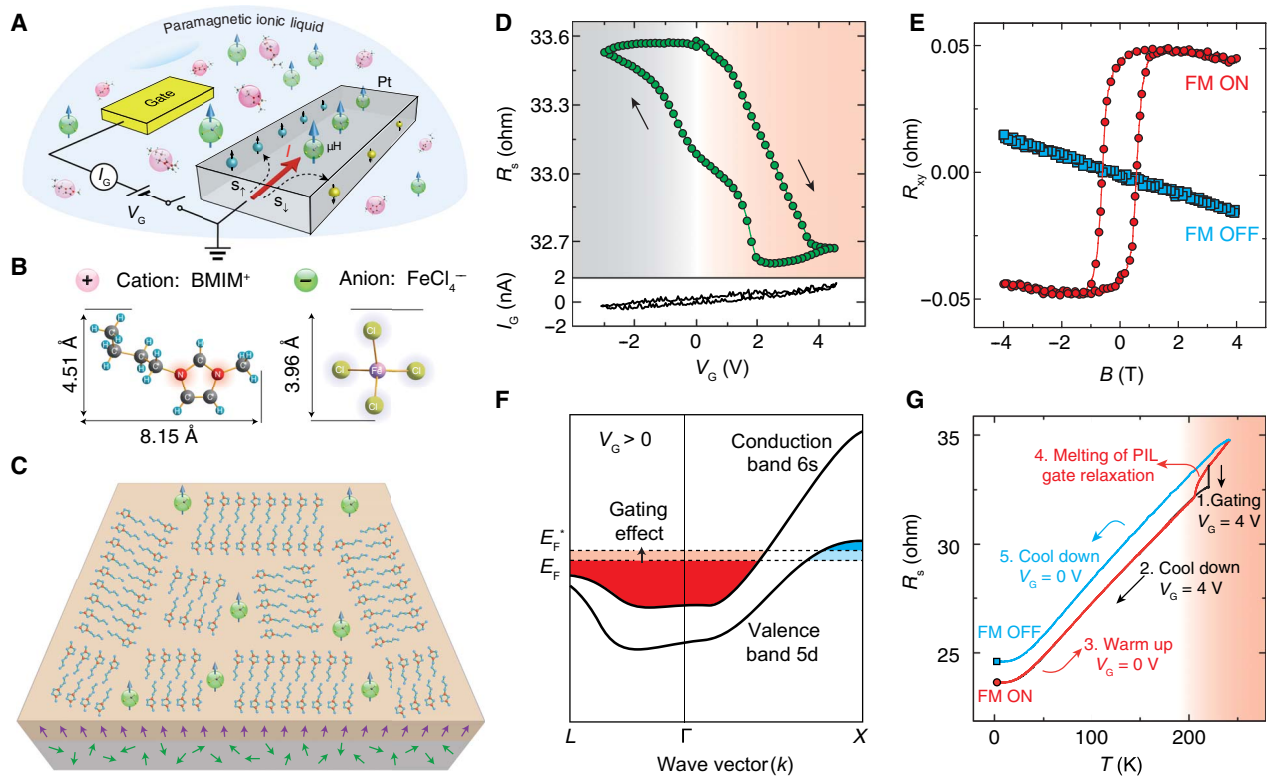

**Fig. 1. Inducing ferromagnetism in Pt films by PIL gating.** (**A**) Schematic diagram of PIL gating and transport measurement setup. The PIL is biased between a gold side gate electrode and the Pt channel. The cations and anions are illustrated as pink and green balls, respectively. (**B**) Composition of a typical PIL used in the measurement: $BMIM^+$ (cation) and $FeCl_4^-$ (anion). The size of the ions is determined from the single-crystal x-ray diffraction at 100 K. (**C**) Cations are driven toward the surface of Pt, forming ordered domain patterns after applying positive $V_G$, which leaves the domain interstitials filled with anions (26). The purple and green arrows denote the spin configuration of the surface and bulk Pt after PIL gating. (**D**) $V_G$ dependence of $R_s$ and corresponding $I_G$ of device A ($t = 8.0$ nm, $-3$ V $\leq V_G \leq 4$ V, and sweeping rate of 50 mV/s). The reversible $V_G$ dependence of $R_s$ with linear $I_G$ profile showing no signature of redox reaction indicates the electrostatic nature of gating. (**E**) $R_{xy}$ measured at 2 K with/without applying $V_G$ in accordance with the gating procedures in (G). The FM OFF and ON denote the linear Hall effect and the AHE (with hysteresis loop) observed in $R_{xy}$ corresponding to the PM and FM states, respectively. (**F**) Field effect tuning of Fermi level $E_F$ in the simplified band structure of Pt, where the Fermi surface of pristine Pt is composed of 5d and 6s states at X and Γ points of the Brillouin zone, respectively. Applying positive $V_G$ lifts the $E_F$ (to $E_F^*$), which changes the ratio between $n_e$ and $n_h$ of 6s and 5d band, respectively. (**G**) Magnetic states of Pt accompanied with five consecutive PIL gating procedures. AHE can be induced after cooling down with $V_G = 4$ V (E, red) corresponding to the FM ON state. Releasing the $V_G$ at 2 K cannot remove the gating effect from the frozen ions. The $R_s$ measured by warming the device from 2 to 260 K with $V_G = 0$ V (red) coincides with the cooling down curve (black) until 200 K. The melt of PIL at higher temperature releases the gating effect; hence, the $R_s$ gradually increases with the redistribution of the accumulated ions. At 220 K, the $R_s$ (red) shows identical value as the state before gating (black), indicating a repeatable gating process. Finally, cooling down the device from 260 to 2 K with $V_G = 0$ V (blue) shows linear Hall effect $R_{xy}$ (E, blue dots) corresponding to the FM OFF state.

is consistent with other reports of the electric-field tuning of magnetic moments in systems with Stoner enhancement, where a positive $V_G$ increases magnetic parameters such as saturation magnetization ($M_s$), coercivity ($H_c$), and $T_C$ (3, 18). This discrepancy between the experimental results and theoretical prediction suggests that the number of 5d electrons of Pt decreases when an electric field is applied in the direction of increasing the total number of electrons, mostly from the s band (24).

As reported in many other material systems (11, 12), ionic gating can cause a significant change in carrier density due to its strong field effect. Here, the apparent carrier density measured by the Hall effect $n_{Hall}$ (for example, extrapolated from $R_{xy}/B$) at 5 K significantly increases from $1.68 \times 10^{17}$ cm$^{-2}$ to $3.24 \times 10^{17}$ cm$^{-2}$ by applying $V_G = 4$ V (Fig. 1D). Because the actual change of carrier density is caused by the formation of electric double layer, the upper bound of the doping concentration can be simply determined by counting the number of ions accumulated at the channel surface. Direct imaging of the ion-gated Au surface by scanning tunneling microscopy (25) shows that the maximum induced carrier density is limited to $\sim 5 \times 10^{14}$ cm$^{-2}$ (Fig. 1C). Therefore, the large discrepancy between the carrier density estimated from the surface ion concentration and the Hall effect indicates that $n_{Hall}$ may not be a suitable indicator to quantify the actual change of carriers.

Despite the quantitative difference, large $\Delta n_{Hall}$ implies substantial change of $E_F$, which is composed of the 6s electron–like and 5d hole–like pockets in open and nearly closed Fermi surfaces, respectively (Fig. 1F) (26–28). Positive $V_G$ lifts $E_F$, accompanied by the increase of 6s electrons ($n_e$) and the decrease of 5d holes ($n_h$). Because of their opposite Hall coefficients, changes of $n_e$ and $n_h$ contribute destructively to the transverse resistance $R_{xy}$ (29), causing seemingly large $\Delta n_{Hall}$. Because of the elevated $E_F$ at $V_G > 0$, the conductivity improves by having more 6s electrons with higher mobility. The increase of $E_F$ also reduces the available empty 5d states, causing less s-d scattering (30). Considering the intrinsically large carrier density in Pt and comparably small change of carriers caused by the field effect, observing large $\Delta R_s$ reflects the influence mainly by the reduced ratio of $n_h/n_e$. The FM state can be switched ON and OFF by following different sequences of PIL gating, as shown in Fig. 1G. The coincidence between the appearance of AHE and the decrease of $R_s$ indicates a close relationship between the emergence of ferromagnetism and PIL gating.

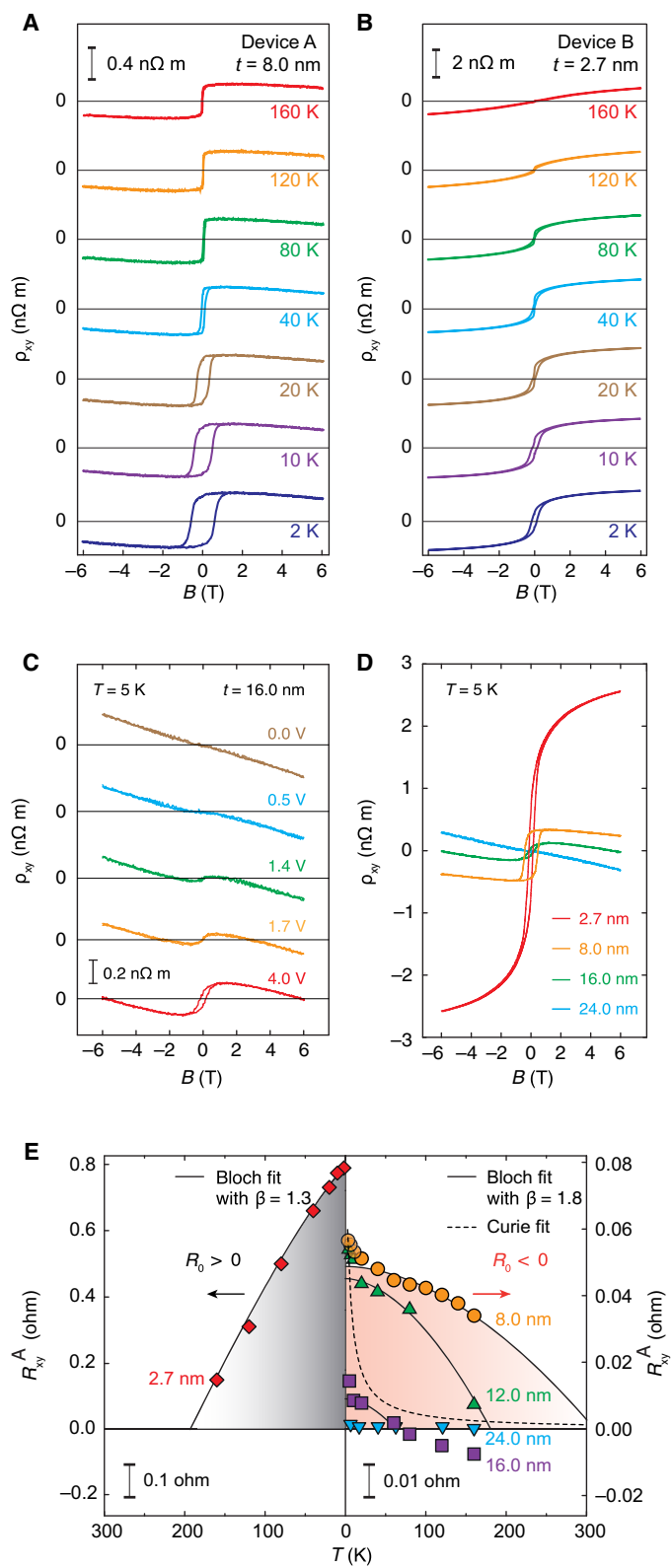

**Fig. 2. Electrical transport of induced FM states after PIL gating.** (**A** and **B**) Temperature variation of Hall resistivity $\rho_{xy}$ for devices A ($t = 8.0$ nm) and B ($t = 2.7$ nm). (**C**) $V_G$ variation of Hall resistivity $\rho_{xy}$ for Pt film with $t = 16$ nm. (**D**) Thickness variation of the Hall resistivity $\rho_{xy}$ for several Pt films. (**E**) Phase diagram of $M_s$ (shown as $R_{xy}^A$) versus temperature for Pt films with different thicknesses.

## The thickness dependence of AHE and magnetic phase diagram

For PIL-gated Pt films with two thicknesses $t = 8.0$ nm (device A) and 2.7 nm (device B), the temperature dependence of the Hall effect shows the occurrence of ferromagnetism manifested as AHE with clear hysteresis loops (Fig. 2, A and B). Empirically, the Hall resistivity in ferromagnets can be described by $\rho_{xy} = \rho_{xy}^0 + \rho_{xy}^A = R_0 B + R_A \mu_0 M$, where $B$ is the magnetic field, $M$ is the magnetization, and $R_0$ and $R_A$ are the ordinary and anomalous Hall coefficients, respectively. Compared with device A, the anomalous Hall term $\rho_{xy}^A$ in device B is one order of magnitude larger, suggesting stronger $M$ in thinner films. It is worth noting that the strength of the induced FM states is closely linked to the $V_G$. As shown in Fig. 2C, the AHE measured for the same device at 5 K shows a gradual decrease of the $\rho_{xy}^A$ and $H_c$ after being cooled down with different $V_G$ biases reduced from $V_G = 4$ V, which is consistent with the increase of $R_0$ demonstrated in Fig. 1F. Because of the strong screening in Pt, the field-induced modulation is confined beneath the surface within a depth of a few angstroms. The large hysteresis loop under the out-of-plane $B$ field implies strong perpendicular magnetic anisotropy (PMA). Therefore, the induced FM state mimics the ideal two-dimensional itinerant Ising ferromagnetism. Magnetic films with PMA are highly pursued in spintronics, acting as the exchange bias layers in giant magnetoresistance devices (31) and magnetic tunnel junctions (32). The easily accessible PMA in PIL-gated Pt is technically favorable because of its electrical tunability.

It is well known that the induced FM state, demonstrated by the emerging hysteresis loop in AHE, should also leave traces in the longitudinal resistance. Surprisingly, the magnetoresistance (MR) shows a very weak signature of hysteresis loop, which is distinguishable only in the thinnest sample ($t = 2.7$ nm), when the influence of bulk Pt is minimal (fig. S5A). On the other hand, clear correspondence between the hysteretic MR and AHE was exhibited in PIL-gated palladium (Pd) film (fig. S6), which has an electronic structure very similar to Pt. The exact reason for this peculiarly small hysteresis in PIL-gated Pt might be related to the details of magnetic domain switching, where the $R_s$ depends on the conductivity of domain walls formed during the magnetization reversal. For instance, anomalous change of MR has been observed in magnetically doped topological insulators (33), where more conductive hysteresis loops were observed in contrast to the conventional, more resistive loops.

As shown in Fig. 2D, the film thickness dependence of $\rho_{xy}$ indicates that the $\rho_{xy}^A$ increases with the decrease of $t$. For films with $t > 24.0$ nm, the AHE signal cannot be distinguished from the linear $\rho_{xy}$ because the short screening length isolates the gating effect from affecting the bulk of gated Pt films, which remains PM with linear Hall response. With the increase of $t$, the enlarging bulk contribution acts as a short-circuit channel bypassing the topmost FM layer, resulting in the reduction of $\rho_{xy}^A$. Despite the larger $\rho_{xy}^A$ in thinner films, the largest $H_c$ was observed for $t = 8.0$ nm. Because the $H_c$ is closely related to the formation of magnetic domains in different film morphologies, the optimization of the density and rigidity of domain walls at this intermediate thickness might give rise to the largest $H_c$.

The FM state is induced in Pt by the formation of an electric double layer, where the thickness is temperature-independent and expected to be a few angstrom thick due to the strong screening effect in a typical metal. We determine the $R_{xy}^A$ by extracting the linear part of $R_{xy}$ under high $B$ field. The temperature dependence of the $M_s$ (that is proportional to the $R_{xy}^A$) can be described by the Bloch equation $M_s(T) = M_s(0)(1 - CT^\beta)$, where $M_s(0)$ is the saturation magnetization at zero temperature and $C$ is a temperature-independent constant. The fittings for the high-temperature region ($T > 40$ K) yield $\beta = 1.3$ and 1.8 for the thinnest ($t = 2.7$ nm) and thicker films ($t = 8.0$, 12.0, and 16.0 nm), respectively (Fig. 2E). The similar $\beta$ and resembling behavior of $R_{xy}^A$ indicate that all FM samples with different thicknesses are likely originated from the same type of magnetization. In general, $T_C$ decreases with the increase of $t$. For sample A ($t = 8.0$ nm) optimized for the largest $H_c$, the extrapolated $T_C$ is even above the room temperature (300 K). The $T_m$ of BMIM[FeCl$_4$] limits the upper-bound temperature of the induced FM state. Alternatively, choosing PILs with higher $T_m$ might enable room temperature FM switching. At low temperature (<40 K), the $M_s$ deviates from the Bloch equation, exhibiting an upturn in accordance with the $1/T$ dependence. This behavior signals the interaction between Pt and the PM FeCl$_4^-$ anions, whose PM magnetization significantly increases at low temperature, obeying Curie's law. It is worth noting that applying the same gating protocol to the identical Pt films using a conventional IL (fig. S7) shows no ferromagnetism despite the increased electrical conductivity (19). This clear difference indicates the importance of PIL in inducing the FM state in Pt.

### The coexistence of Kondo effect and surface FM states

Depositing magnetic molecules on a metal film provides the ingredients essential to induce the Kondo effect, which has been observed in systems where a monolayer of a metal complex with unpaired spins is deposited on a gold (Au) film (34, 35). The ionic gating has also been applied to study the Kondo effect (36, 37). After cooling down the devices with positive $V_G$ at which the $R_s$ approaches saturation (Fig. 3A), systematically, we observed the $R_s$ upturns below 20 K for all FM films with different thicknesses (Fig. 3B). The saturation of $R_s$ toward zero temperature away from the linear dependence on $\ln(T)$ (inset of Fig. 3B) suggests the coexistence of the Kondo effect.

To quantitatively extract key parameters such as the Kondo resistance ($R_K$) and the Kondo temperature ($T_K$), we performed a numerical renormalization group (NRG) analysis (38), which normalizes the Kondo contribution to the transport by a universal parameter $T/T_K$. As shown in Fig. 3C, the $R_K(T)/R_K(0)$ dependences obtained for all measured films collapse into a single fitting curve as a function of $T/T_K$, which verifies the Kondo effect as the origin of the observed $R_s$ upturns. As an example, a typical $R_s$ upturn for $t$ = 2.7 nm can be fitted well by the Kondo scenario (inset of Fig. 3C). When $T > T_K$, the increasing deviation from the NRG fitting is mainly attributed to the stronger phonon scattering.

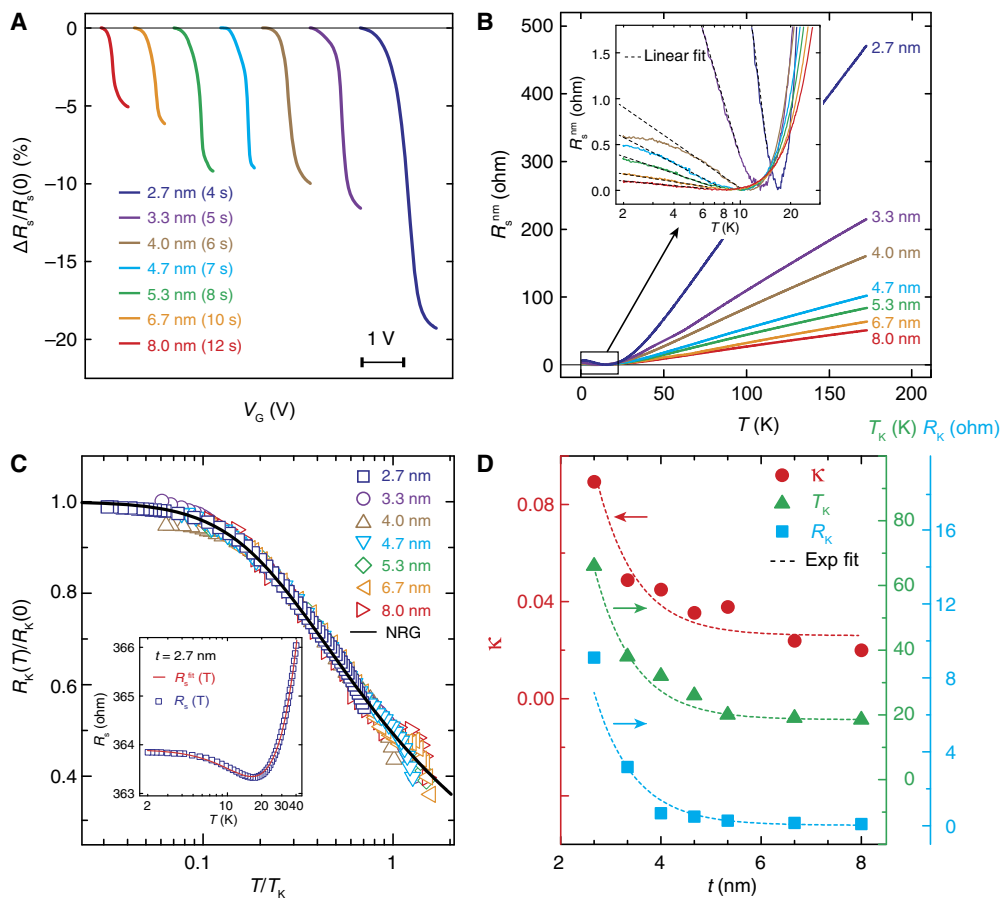

**Fig. 3. Gate-induced Kondo effect in Pt films with different thicknesses.** (**A**) $V_G$ dependence of the $R_s$ ($T = 220$ K) for a series of Pt films with different thicknesses (proportional to the sputtering time) showing the regime where the $R_s$ rapidly decreases. (**B**) Temperature dependence of the $R_s$ ($V_G = 4$ V) showing the Kondo effect in Pt films with different thicknesses. The $R_s$ curves are normalized to the $R_s$ measured at 150 K, according to $R_s^{nm}(T) = \frac{R_s(T) - R_s^{min}}{R_s(150\,K) - R_s^{min}} \times R_s(150\,K)$, where $R_s^{min}$ is the minimum $R_s$ of each curve. The inset shows the low temperature region, where the $R_s^{nm}$ increases logarithmically with the decrease of $T$ before reaching saturation. (**C**) Universal Kondo behavior observed for all gated Pt films ($V_G = 4$ V) showing that the normalized Kondo resistance $R_K(T)/R_K(0\,K)$ versus the reduced temperature $T/T_K$ collapses into a single line, which can be described by the NRG method (black line). The inset shows an upturn in the temperature dependence of the $R_s$ ($t = 2.7$ nm) observed from 2 to 30 K. The red line is the fitting from the NRG equation: $R_s(K) = R_0 + aT^b + R_K(T/T_K)$, where $R_0$ is the residual resistance, and $a$ and $b$ are temperature-independent coefficients. The empirical function $R_K(T/T_K)$ can be further expanded into $R_K(T/T_K) = R_K(0\,K)\left(\frac{T_K^{'2}}{T^2 + T_K^{'2}}\right)^s$, where $R_K(0\,K)$ is the Kondo resistance at zero temperature, $T_K^{'} = T_K/(2^{1/s} - 1)^{1/2}$, and the parameter $s$ is fixed to 0.225. (**D**) Thickness dependence of Kondo parameters $R_K$ and $T_K$ and effective thickness ratio $\kappa$. The dashed lines represent the exponential decay as a function of the film thickness $t$, where the decay constant $c_0 = 0.8$ is the same for all fittings.

Looking at the phenomenon of surface magnetization, the transport phenomena can be described by a simple model composed of two parallel conducting channels, where the ratio between the gate-tuned top layer and the total film thickness is denoted as $\kappa = t_{tp}/t$ (table S1). The estimated thickness of FM Pt is roughly equal to the topmost atomic layer $t_{tp}$ (~2 Å). Both the gating-related parameter $\kappa$ and Kondo parameters $R_K$ and $T_K$ decrease exponentially in the same trend as a function of $t$, indicating a close relationship between PIL gating and the induced Kondo effect (Fig. 3D). The narrowly confined FM state is in contrast with the much longer Kondo cloud length $l_K$ (~5 nm) estimated by $l_K = \sqrt{\hbar D/k_B T_K}$, where $k_B$ and $\hbar$ are the Boltzmann and reduced Planck's constant, respectively. Here, the diffusion constant $D$ is given by $D = \frac{1}{3} v_F l_e$, where $v_F$ is the Fermi velocity and $l_e$ is the elastic mean free path (39).

## DISCUSSION

The origin of the AHE can be analyzed by scaling the anomalous Hall conductivity $\sigma_{AH} = \rho_{AH}/(\rho_{xx}^2 + \rho_{xy}^2)$ as a function of the longitudinal conductivity $\sigma_{xx} = \rho_{xx}/(\rho_{xx}^2 + \rho_{xy}^2)$ in a power law form as $\sigma_{AH} \propto \sigma_{xx}^\gamma$ (40). Figure 4 shows the scaling diagram of a comprehensive set of AHE data measured in different families of magnetic materials. In our experiment, the scaling of $\sigma_{AH}$ is different from the asymmetric skew scattering from impurities ($\sigma_{AH} \propto \sigma_{xx}$) (41) and the side jump mechanism ($\sigma_{AH} \propto$ const.) (42). The fact that $\sigma_{AH}$ appears to be independent of $\sigma_{xx}$ implies that the induced ferromagnetism in Pt is of an intrinsic origin (43, 44). For the PIL-induced FM state of Pt, the temperature dependence of $\sigma_{AH}$ strongly relies on the applied $V_G$ biases (Fig. 2C), which affects not only $\sigma_{xx}$ but also $\sigma_{AH}$. The $\sigma_{xx}$ of gated Pt film shows weak temperature dependence, whereas the $\sigma_{AH}$ decreases significantly. These factors result in a large scaling exponent compared to other materials with $\gamma = 1$ and 2, which is also different from the behaviors observed in conventional itinerant ferromagnets (45). When Pt is deposited on a FM insulator, $Y_3Fe_5O_{12}$ (YIG), the AHE can be induced by the magnetic proximity effect (MPE) (46), showing similar sign reversal of the $R_0$ for $t < 3.0$ nm (47). We attributed this to the significant change of band structure at reduced dimensionality. In contrast to YIG, the strong Coulomb repulsion between anions in BMIM[FeCl$_4$] sets a large interionic distance, prohibiting FM exchange interaction. Hence, the PIL used in our gating remains PM down to 2 K (fig. S2), which firmly excludes MPE as the origin of the induced FM state.

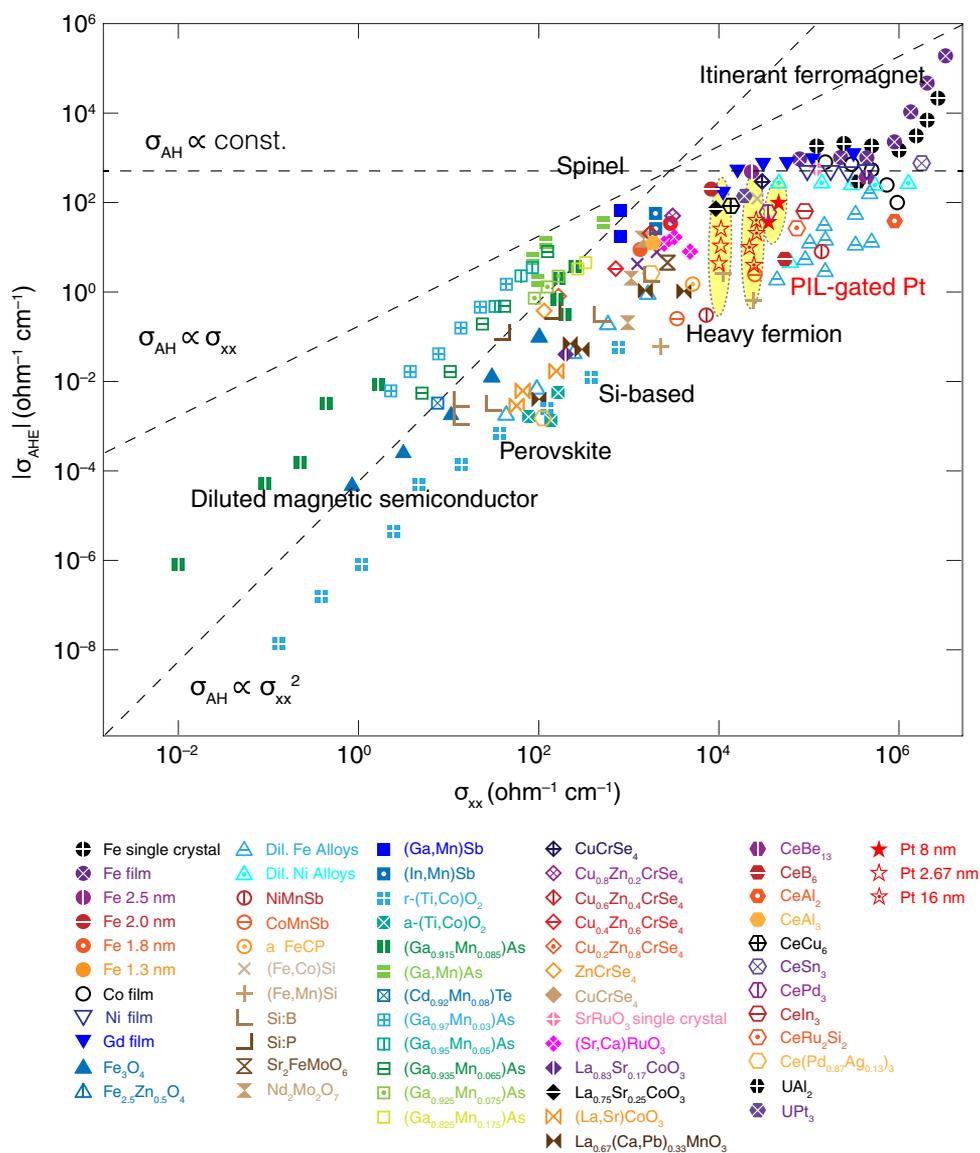

**Fig. 4. Scaling diagram of AHE for PIL-gated Pt and various other magnetic systems.** The AHE measured in PIL-gated Pt samples (shaded by yellow) are plotted together with other magnetic conductors reported previously (53–56), where each type of marker represents one material system. The $\sigma_{AH}$ measured for the FM state of Pt shows weak dependence on the $\sigma_{xx}$.

Itinerant ferromagnetism is generally considered to be detrimental to the Kondo effect, whereas coexistence of FM and Kondo states has been observed in atomic contacts of pure itinerant ferromagnets (48) and heavy-fermion metals (49), where the coexistence is due to the local moments formed in a reduced coordinating environment and the itinerant-localized duality of f electrons, respectively. Although a comprehensive physical picture of the coexisting FM and Kondo states requires further study, the emerging ferromagnetism at the PIL/Pt interface likely originates from the d-d interaction between the 5d electrons perturbed by the field effect and the local magnetic moment of the metal halide anions. Although the FM state is confined at the PIL/Pt interface due to the strong screening, the spatially separated Kondo scattering extends deeper into the film, causing coexistence as two parallel channels.

To substantiate our observation that the induced FM state and Kondo effect in Pt films are caused by PIL gating, we performed a series of control experiments. By successively gating using different ionic media (IL-PIL-IL-PIL) on the same device (fig. S9), we firmly demonstrate the indispensable role of the PIL in inducing ferromagnetism. Compared with PIL-gated samples, the Pt film of the same thickness ($t = 8.0$ nm) with intentional Fe contamination underneath (nominal thickness, 0.2 nm) shows negligible AHE and $H_c$ in the absence of the Kondo effect (fig. S5B). Moreover, the PIL-gated Au film maintains a PM state (fig. S10) because of its fully filled 5d band, whereas the Pd film yields a similar FM state after PIL gating (fig. S6). It is worth noting that although exclusive correlation has been demonstrated between PIL gating and the induced FM state, the exact carrier doping mechanism might not be limited to the electrostatic effect. Consistent with the fully reversible and repeatable behaviors shown in the control experiments (fig. S9), the reversible, electrochemically induced phase transformation that is stabilized by gating can be the alternative doping mechanism (50). Therefore, probing the redox states of the gated surface would help in resolving the exact doping mechanism (51). Our present results of inducing FM states reveal the universality of PIL gating in switching magnetism, which will benefit the development of spintronic devices that can simultaneously control charge and spin degrees of freedom.

## MATERIALS AND METHODS
### Synthesis of PILs
A series of PILs (fig. S1A) were synthesized by mixing stoichiometric amounts of solid organic-based halides and anhydrous transition metal halides (Sigma-Aldrich) inside a $N_2$-filled glove box (52). The ratios of the two components were calculated based on the selection of the corresponding metal chlorides, taking the oxidation states of the metal ions (for example, 3+ and 2+) and the coordination number of the corresponding metal complex (for example, MR$_4$ and MR$_6$) into account. For example, to synthesize BMIM[FeCl$_4$], which was later chosen for all transport measurements, BMIM[Cl]/anhydrous FeCl$_3$ precursors with a molar ratio of 1.05:1 were chosen to ensure that all Fe elements contained in the PIL could stay in the form of FeCl$_4^-$. The mixture was dispersed in dichloromethane by stirring overnight at room temperature to form the PIL. The dichloromethane solvent and other volatile residual impurities in the PIL were removed by a rotary vacuum evaporator ($P < 10^{-3}$ mbar). The as-prepared BMIM[FeCl$_4$] shows a strong response to a magnet even at room temperature (fig. S1B).

### Single-crystal x-ray diffraction
The crystal structures of PIL were determined by x-ray diffraction using a Bruker D8 Venture diffractometer equipped with a monochromator (Triumph) and an area detector (Photon 100). We used Mo $K_\alpha$ radiation and carried out the measurement at 100 K to minimize the thermal vibrations. The PIL crystals were picked up using nylon loop with cryo-oil and cooled using a nitrogen flow (Cryostream Plus, Oxford Cryosystems). The diffraction data were processed using the Bruker Apex II software. The crystal structures were solved and refined using the SHELXTL software.

### Device fabrication
Transistor devices used for PIL gating were all fabricated by standard microfabrication. Using electron beam lithography, we defined the Hall bar at $3 \times 7$ μm$^2$. All metal channels (Pt, Pd, and Au) were prepared by dc magnetron sputtering (Kurt J. Lesker) after pumping the chamber below $1.0 \times 10^{-8}$ mbar. Sputtering powers (50 to 200 W) and duration (2 to 12 s) were optimized to prepare films with different thicknesses. Separately, contact electrodes comprising bilayer Ti/Au (5/70 nm) were deposited onto the patterned Hall bars using e-beam evaporation (FC-2000, Temescal) below $1.0 \times 10^{-6}$ mbar. Afterward, an Al$_2$O$_3$ isolation layer (30 nm) was deposited to cover contact electrodes, limiting the gating effect to only the exposed channel surface.

### Electrical transport measurement
Low-temperature electrical transports were measured in a helium cryostat (Physical Property Measurement System, Quantum Design) under out-of-plane magnetic fields up to 6 T. All transport properties were measured by two lock-in amplifiers (SR830, Stanford Research) using a constant ac current excitation of 50 μA at 13.367 Hz. The voltage bias on BMIM[FeCl$_4$] (the PIL used in all gating experiments) was applied by a source measure unit (model 2450, Keithley).

### Magnetic property characterization
The magnetization curves of BMIM[FeCl$_4$] were measured in a SQUID magnetometer (MPMS XL-7, Quantum Design) up to 7 T. The temperature dependence of magnetic susceptibility was measured using zero-field cooling (ZFC) and field cooling (FC) methods. In the ZFC method, the sample was first cooled down to 2 K in zero field and measured during warming up in a field of 100 Oe. In the FC method, the measurement procedure was identical except that PIL was first cooled down in a field of 2 T.

## SUPPLEMENTARY MATERIALS
Supplementary material for this article is available at http://advances.sciencemag.org/cgi/content/full/4/4/eaar2030/DC1

section S1. Magnetization and magnetic susceptibility measurement section S2. Two-channel model for calculating Kondo effect measurement
section S3. Charge accumulation dynamics
section S4. Transport properties measured for PIL gating with $V_G < 0$
section S5. Gate dependence of AHE
section S6. Fe impurity–doped Pt films
section S7. Transport properties of PIL-gated palladium film
section S8. MR of pristine, conventional IL-gated, and PIL-gated Pt films
section S9. Gating cycles with sequential switch of the gating media between PIL and conventional IL
section S10. Optical and atomic force microscopy images of a typical Pt sample
section S11. Transport properties of PIL-gated gold film
fig. S1. Various kinds of PILs and their response to magnets.
fig. S2. Temperature variation of the magnetic properties of BMIM[FeCl$_4$].
fig. S3. Chronoamperometry measurement of PIL gating.
fig. S4. The transport measurement of PIL-gated Pt film under negative $V_G$.
fig. S5. Comparison between Fe impurities contaminated and PIL-gated Pt thin films.
fig. S6. Low-temperature ($T$ = 5 K) electrical transport of pristine and PIL-gated Pd thin films.
fig. S7. Comparison between low temperature ($T$ < 40 K) MRs and $B$-dependent $R_s$ under various gating conditions.
fig. S8. Optical and atomic force microscopy images of a typical Pt sample.
fig. S9. Gating cycles by switching the gating media between nonmagnetic IL and PIL on the same Pt film ($t$ = 12.0 nm).
fig. S10. Transport properties of PIL-gated Au thin film.
table S1. Fitting parameters of the Kondo effect for films with different thicknesses.
References (57–62)

**Acknowledgments:** We thank B. J. van Wees and G. E. W. Bauer for fruitful discussion. **Funding:** The work was supported by the European Research Council (consolidator grant no. 648855 Ig-QPD) and Dutch national facility NanoLabNL. L.L. was supported by the Ubbo Emmius scholarship of University of Groningen. **Author contributions:** J.Y. and L.L. conceived the experiments. L.L. and W.T. fabricated the devices. L.L. designed and carried out the transport experiments, with help from Q.C. and J.L. J.S., L.L., G.R.B., and T.T.M.P. measured the crystallographic and magnetic properties. J.Y., L.L., and T.T.M.P. constructed the theory. All authors were involved in the data analysis. L.L. and J.Y. wrote the manuscript, with the help of co-authors. **Competing interests:** The authors declare that they have no competing interests. **Data and materials availability:** All data needed to evaluate the conclusions in the paper are present in the paper and/or the Supplementary Materials. Additional data related to this paper may be requested from the authors.

Submitted 12 October 2017
Accepted 16 February 2018
Published 6 April 2018
10.1126/sciadv.aar2030

**Citation:** L. Liang, Q. Chen, J. Lu, W. Talsma, J. Shan, G. R. Blake, T. T. M. Palstra, J. Ye, Inducing ferromagnetism and Kondo effect in platinum by paramagnetic ionic gating. *Sci. Adv.* **4**, eaar2030 (2018).

# Science Advances

## Inducing ferromagnetism and Kondo effect in platinum by paramagnetic ionic gating

Lei Liang, Qihong Chen, Jianming Lu, Wytse Talsma, Juan Shan, Graeme R. Blake, Thomas T. M. Palstra and Jianting Ye